\documentclass[12pt]{emulateapj}

\usepackage{natbib}

\newcommand{\teff}{${T}_{\mathrm{eff}}$}
\newcommand{\logg}{$\log{g}$}
\newcommand{\msun}{$M_{\odot}$}

\newcommand{\kms}{km s$^{-1}$}
\newcommand{\muhz}{$\mu$Hz}
\newcommand{\Te}{${T}_{\mathrm{eff}}$}

\shorttitle{Discovery of the Second and Third Pulsating He-Core WDs}
\shortauthors{Hermes et al.}

\begin{document}

\title{DISCOVERY OF PULSATIONS, INCLUDING POSSIBLE PRESSURE MODES, IN TWO NEW EXTREMELY \\ LOW MASS, HE-CORE WHITE DWARFS}
\author{J. J. Hermes\altaffilmark{1,2}, M. H. Montgomery\altaffilmark{1,2}, D. E. Winget\altaffilmark{1,2}, Warren R. Brown\altaffilmark{3}, A. Gianninas\altaffilmark{4}, \\ Mukremin Kilic\altaffilmark{4}, Scott J. Kenyon\altaffilmark{3}, Keaton J. Bell\altaffilmark{1,2}, and Samuel T. Harrold\altaffilmark{1,2}}

\altaffiltext{1}{Department of Astronomy, University of Texas at Austin, Austin, TX\,-\,78712, USA}
\altaffiltext{2}{McDonald Observatory, Fort Davis, TX\,-\,79734, USA}
\altaffiltext{3}{Smithsonian Astrophysical Observatory, 60 Garden St, Cambridge, MA\,-\,02138, USA}
\altaffiltext{4}{Homer L. Dodge Department of Physics and Astronomy, University of Oklahoma, 440 W. Brooks St., Norman, OK\,-\,73019, USA}

\email{jjhermes@astro.as.utexas.edu}


\begin{abstract}

We report the discovery of the second and third pulsating extremely low mass white dwarfs (WDs), SDSS~J111215.82+111745.0 (hereafter J1112) and SDSS~J151826.68+065813.2 (hereafter J1518). Both have masses $<$ 0.25 \msun\ and effective temperatures below $10,000$ K, establishing these putatively He-core WDs as a cooler class of pulsating hydrogen-atmosphere WDs (DAVs, or ZZ Ceti stars). The short-period pulsations evidenced in the light curve of J1112 may also represent the first observation of acoustic ($p$-mode) pulsations in any WD, which provide an exciting opportunity to probe this WD in a complimentary way compared to the long-period $g$-modes also present. J1112 is a \teff\ $=9590\pm140$ K and \logg\ $=6.36\pm0.06$ WD. The star displays sinusoidal variability at five distinct periodicities between $1792-2855$ s. In this star we also see short-period variability, strongest at 134.3 s, well short of expected $g$-modes for such a low-mass WD. The other new pulsating WD, J1518, is a \teff\ $=9900\pm140$ K and \logg\ $=6.80\pm0.05$ WD. The light curve of J1518 is highly non-sinusoidal, with at least seven significant periods between $1335-3848$ s. Consistent with the expectation that ELM WDs must be formed in binaries, these two new pulsating He-core WDs, in addition to the prototype SDSS~J184037.78+642312.3, have close companions. However, the observed variability is inconsistent with tidally induced pulsations and is so far best explained by the same hydrogen partial-ionization driving mechanism at work in classic C/O-core ZZ Ceti stars.

\end{abstract}



\section{Introduction}

There are many pulsation instability strips on the Hertzsprung-Russell diagram, including the DAV (or ZZ Ceti) instability strip, driven by a hydrogen partial ionization zone in the cool, hydrogen atmosphere (DA) white dwarfs (WDs). Seismology using the non-radial gravity-mode ($g$-mode) pulsations of DAVs has enabled us to constrain the mass, core and envelope composition, rotation rate, and the behavior of convection in these objects (see reviews by \citealt{WinKep08} and \citealt{FontBrass08}). However, the roughly 150 DAVs known to date have masses between $0.5-1.1$ \msun, implying that they all likely contain C/O-cores. Low-mass, He-core WDs are likely to pulsate as well, including extremely low-mass (ELM, $\leq 0.25$ \msun) WDs \citep{Steinfadt10}.

ELM WDs are the byproducts of binary evolution. Mass loss during at least one common-envelope phase has precluded the ELM WD from igniting He in its core, and the result is an underweight WD with a core devoid of C/O. These WDs have been known for some time as companions to pulsars, but recently their numbers have grown dramatically as a result of the ELM Survey, a targeted spectroscopic search for ELM WDs \citep{BrownELMi,KilicELMii,BrownELMiii,KilicELMiv}.

Leveraging the numerous new low-mass WDs found by the ELM Survey, the first putatively He-core ELM WD (SDSS J184037.38+642312.3, hereafter J1840) was recently discovered to pulsate \citep{HermesJ1840}. This $\sim$0.17 \msun\ WD varies at a dominant period of roughly 4698 s, with a high-amplitude ($>5$\%), non-sinusoidal pulse shape. The discovery of pulsating ELM WDs provides the first chance to apply the tools of asteroseismology to these low-mass, presumed He-core WDs; we have an exciting opportunity to probe the internal structure of these exotic compact objects \citep{Corsico12,VanGrootel13}.

Here we announce the discovery of pulsations in two new ELM WDs: SDSS J111215.82+111745.0 (hereafter J1112) and SDSS J151826.68+065813.2 (hereafter J1518). These objects bring to three the number of pulsating ELM WDs known, further populating the empirical instability strip of ELM WDs. These objects belong to an extension of the instability strip from the C/O-core DAVs, and are most likely driven by the same mechanism acting in the classical ZZ Ceti stars.

In Section~\ref{sec:J1112} we detail our discovery of pulsations in J1112. We outline our new spectroscopic and photometric observations of this ELM WD, and provide a robust identification of its highest-amplitude periodicities. We also detail evidence for short-period variability which may be evidence of acoustic ($p$-mode) pulsations in this ELM WD. In Section~\ref{sec:J1518} we describe the discovery of variability in the ELM WD J1518 and list the significant periods determined from our relatively sparse coverage so far. We conclude with a discussion of these discoveries in Section~\ref{sec:end}.



\section{SDSS J1112+1117}
\label{sec:J1112}

\subsection{Spectroscopic Observations}
\label{sec:J1112spec}

\citet{BrownELMiii} presented a preliminary fit of \teff\ $=9400\pm490$ K and \logg\ $= 5.81\pm0.12$ based on a single spectrum of this $g_0=16.2$ mag\footnote{$g_0 = $ SDSS dereddened $g$-band magnitude} WD from the FLWO 1.5 m telescope using the FAST spectrograph \citep{Fabricant98}. We have obtained an additional 26 spectra using the FLWO 1.5 m telescope and an additional 6 spectra using the Blue Channel Spectrograph \citep{Schmidt89} on the 6.5m MMT. The time-series spectroscopy reveals that J1112 is short-period binary, as described below.

\subsubsection{Atmospheric Parameters}
\label{sec:atm}

We have phased (see Section~\ref{sec:J1112rv}) and co-added our new spectroscopic observations to determine the atmospheric parameters of the primary ELM WD visible in J1112. Our observations cover a wavelength range from $3700-4500$ \AA. The model atmospheres used for this analysis are described at length in \citet{Gianninas11} and employ the new Stark broadening profiles from \citet{TB09}. Models where convective energy transport becomes important are computed using the ML2/$\alpha$ = 0.8 prescription of the mixing-length theory \citep[see][]{tremblay10}. However, since we are dealing with ELM WDs, the model grid needed to be extended down to lower surface gravities. Thus the model grid used in this analysis covers the ranges in \Te\ from 4000 to 30,000 K in steps ranging from 250 to 5000 K and \logg\ from 5.0 to 8.0 in steps of 0.25 dex.

The method used for fitting the observations relies on the so-called spectroscopic technique, described in \citet{Gianninas11} and references therein. The main difference here is that we fit the Balmer lines we observe up to and including H12, as the considerably lower surface gravity means that these higher Balmer lines are still observed. Since the higher Balmer lines are sensitive mostly to \logg\ (see Figure 2 of \citealt{TB09}), the inclusion of these extra lines further constrains our measurement of the surface gravity.

The uncertainty for each parameter is calculated by combining the internal error, which is the dominant source of uncertainty, obtained from the covariance matrix of the fitting algorithm with the external error, obtained from multiple observations of the same object, estimated for DA stars at 1.2\% in \Te\ and 0.038 dex in \logg\ (see \citealt{lbh05} for details). The effect on the uncertainties caused by different values of S/N are included in the internal error.

\begin{figure}[t]
\centering{\includegraphics[width=0.9\columnwidth]{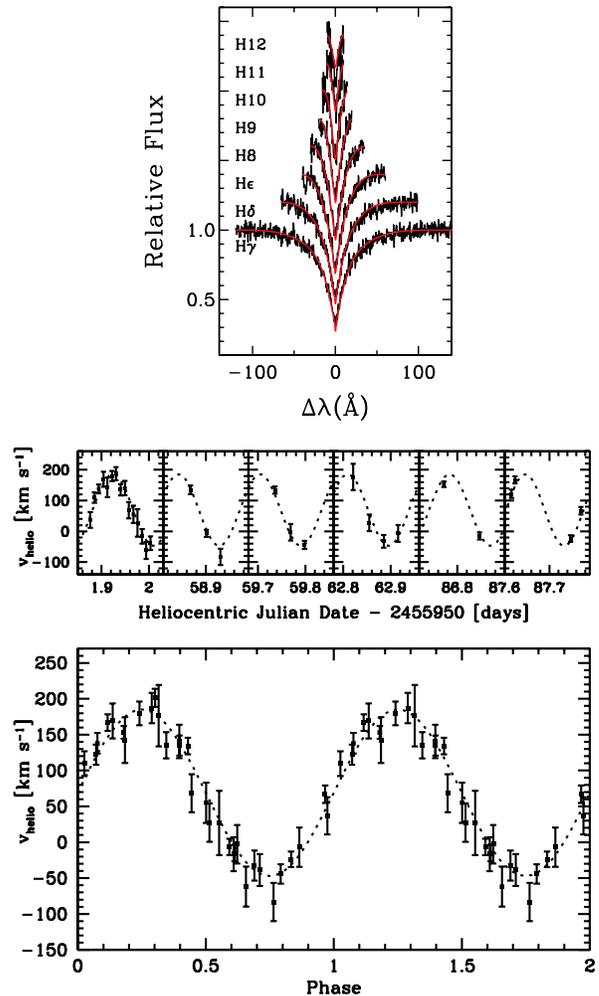}}
\caption{Spectroscopic observations of J1112. The top panel shows the summed and phased spectra, with a model fit to the H$\gamma$--H12 lines of the Balmer series. This model derives the primary parameters in Section~\ref{sec:atm}. The middle panel shows our new radial velocity observations of J1112 over six epochs, and the bottom panel shows those data phased to the orbital period of $4.13952$ hr. \label{fig:J1112rv}}
\end{figure}

Our final fit to the phased and co-added spectra of J1112 is shown in the top panel of Figure~\ref{fig:J1112rv} and yield \teff\ $=9590\pm140$ K and \logg\ $=6.36\pm0.06$. This corresponds to a mass of $\sim$0.17 \msun\ using the He-core models of \citet{Panei07}.

In addition to the Balmer series, the Ca II K line at 3933 \AA\ is also observed in the spectra of J1112. For the purposes of this analysis, we simply exclude the wavelength range where that metal line is present so that it does not affect either the normalization of the individual Balmer lines nor the actual fits themselves. This Ca feature phases with the radial velocity variations of the Balmer lines, and is thus not interstellar but rather in the photosphere of the WD. We will not discuss this Ca feature further in this paper.

\subsubsection{Radial Velocity Observations}
\label{sec:J1112rv}

The bottom two panels of Figure~\ref{fig:J1112rv} show the radial velocity curve of the 32 spectra we have obtained of J1112. We compute the orbital elements of this single-lined binary using the code of \citet{Kenyon86}, which weights each velocity measurements by its associated error. Our spectroscopic observations find that J1112 is in a $4.13952\pm0.00024$ hr orbital period binary with a $K=116.2\pm2.8$ \kms\ radial velocity semi-amplitude.

This yields a mass function of $f = 0.028\pm0.003$ \msun, which constrains the minimum mass of the unseen companion to $M_2 > 0.14$ \msun\ assuming a 0.17 \msun\ primary. The nature of this companion has no direct bearing on the discovery of pulsations in the primary ELM WD, but it may be contribute enough flux to partially contaminate the spectral fits of the primary. It also suggests that unless the system is inclined to less than $30$ degrees (which would happen at random less than 15\% of the time), the companion is most likely another He-core WD with a mass below 0.45 \msun. Similar single-lined ELM WDs with likely He-core WD companions have been found in the ELM Survey (i.e., SDSS J1005+3550 in \citealt{KilicELMiv}).

We have folded our photometric observations on this orbital period and do not see evidence of eclipses, at a limit of 0.5\%. This suggests that the inclination of the system is $i < 80$ degrees, although this does not significantly constrain the nature of the unseen companion.

\subsection{Photometric Observations}
\label{sec:J1112photo}

We obtained high-speed photometric observations of J1112 at the McDonald Observatory over four months, from 2012 January to 2012 April, for a total of more than 70.2 hr of coverage. We used the Argos instrument, a frame-transfer CCD mounted at the prime focus of the 2.1m Otto Struve telescope \citep{Nather04}, to obtain $5-10$ s exposures on J1112; a full journal of observations can be found in Table~\ref{tab:jour}. Observations were obtained through a 2mm BG40 filter to reduce sky noise.

\begin{deluxetable}{llccc}
\tabletypesize{\scriptsize}
\tablecolumns{5}
\tablewidth{0.38\textwidth}
\tablecaption{Journal of photometric observations. \label{tab:jour}}
\tablehead{
\colhead{Run} & \colhead{UT Date} & \colhead{Length} & \colhead{Seeing} & \colhead{Exp.}
\\ \colhead{} & \colhead{} & \colhead{(hr)} & \colhead{(\arcsec)} & \colhead{(s)} }
\startdata 
\multicolumn{5}{c}{\bf SDSS J1112+1117} \\
A2568	&	2012 Jan 21	&	1.9	&	3.3	&	5	\\
A2571	&	2012 Jan 22	&	1.3	&	2.2	&	10	\\
A2574	&	2012 Jan 23	&	0.8	&	2.0	&	5	\\
A2577	&	2012 Jan 24	&	2.2	&	1.3	&	10	\\
A2578	&	2012 Jan 25	&	2.5	&	3.4	&	10	\\
A2581	&	2012 Jan 27	&	6.5	&	3.0	&	5	\\
A2586	&	2012 Jan 29	&	2.6	&	1.7	&	5	\\
A2588	&	2012 Jan 30	&	3.2	&	2.0	&	5	\\
A2590	&	2012 Jan 31	&	3.0	&	1.2	&	5	\\
A2594	&	2012 Feb 1	&	3.5	&	2.1	&	5	\\
A2597	&	2012 Feb 2	&	2.6	&	1.9	&	10	\\
A2600	&	2012 Feb 3	&	2.8	&	1.5	&	5	\\
A2605	&	2012 Feb 15	&	3.9	&	1.6	&	10	\\
A2607	&	2012 Feb 16	&	2.4	&	1.8	&	10	\\
A2611	&	2012 Feb 19	&	5.0	&	2.1	&	10	\\
A2614	&	2012 Feb 20	&	2.9	&	1.8	&	10	\\
A2616	&	2012 Mar 12	&	2.4	&	2.4	&	5	\\
A2618	&	2012 Mar 13	&	2.9	&	1.4	&	10	\\
A2621	&	2012 Mar 14	&	3.6	&	1.4	&	10	\\
A2638	&	2012 Mar 17	&	3.1	&	2.5	&	10	\\
A2641	&	2012 Apr 14	&	1.9	&	3.3	&	5	\\
A2656	&	2012 Apr 19	&	3.7	&	2.0	&	5	\\
A2659	&	2012 Apr 20	&	2.2	&	1.5	&	5	\\
A2662	&	2012 Apr 21	&	1.7	&	1.3	&	5	\\
A2675	&	2012 Apr 24	&	1.7	&	1.1	&	5	\\
\multicolumn{5}{c}{\bf SDSS J1518+0658} \\
A2626	&	2012 Mar 15	&	0.7	&	3.3	&	5	\\
A2634	&	2012 Mar 16	&	2.2	&	2.2	&	10	\\
A2639	&	2012 Mar 17	&	3.4	&	3.4	&	10	\\
A2647	&	2012 Apr 16	&	4.9	&	1.7	&	5	\\
A2650	&	2012 Apr 17	&	1.7	&	2.0	&	5	\\
A2660	&	2012 Apr 20	&	1.7	&	1.2	&	5	\\
A2664	&	2012 Apr 21	&	3.9	&	1.9	&	10	\\
A2673	&	2012 Apr 23	&	0.7	&	1.5	&	5	\\
A2676	&	2012 Apr 24	&	0.4	&	1.6	&	10	\\
A2684	&	2012 Jun 18	&	2.7	&	2.1	&	10	\\
A2686	&	2012 Jun 19	&	2.9	&	1.8	&	10	\\
A2688	&	2012 Jun 20	&	2.3	&	2.4	&	5	\\
A2705	&	2012 Jul 13	&	3.1	&	1.3	&	5
\enddata
\end{deluxetable}

\begin{figure}[t]
\centering{\includegraphics[width=0.95\columnwidth]{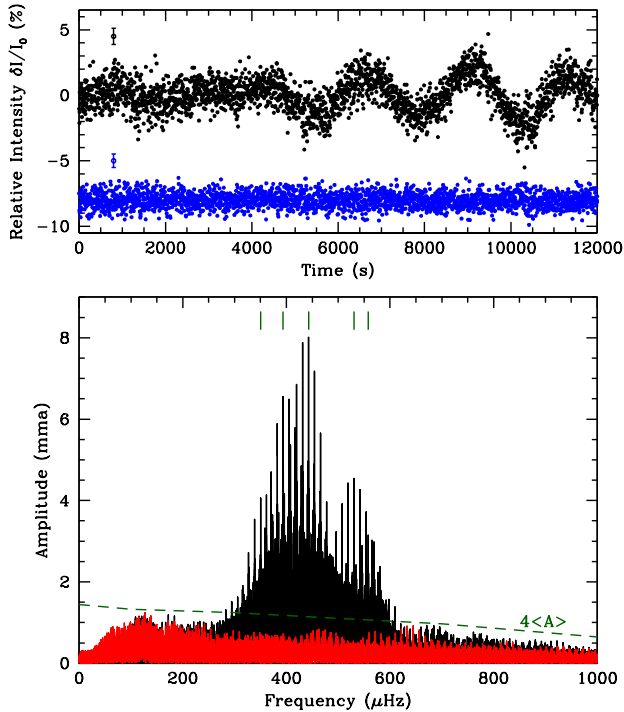}}
\caption{The top panel shows high-speed photometry of J1112 from a representative run on 2012 February 1. The brightest comparison star is shown in blue, offset by $-8$\%. Average point-by-point photometric errors are also shown. The bottom panel shows a Fourier transform of our entire data set to date, some 70.2 hr of observations from 2012~January to 2012~April. We also display in red the Fourier transform of the residuals after prewhitening by the five highest-amplitude periods listed in Table~\ref{tab:J1112freq} and mark those periods with green tick marks at the top of the panel. We mark the 4$\langle {\rm A}\rangle$ significance level as a dashed green line. \label{fig:J1112}}
\end{figure}

We performed weighted, circular, aperture photometry on the calibrated frames using the external IRAF package $\textit{ccd\_hsp}$ written by Antonio Kanaan \citep{Kanaan02}. We divided the sky-subtracted light curves by the brightest comparison star in the field, SDSS~J111211.51+111648.4 ($g=15.0$ mag), to remove transparency variations, and applied a timing correction to each observation to account for the motion of the Earth around the barycenter of the solar system \citep{Stumpff80,Thompson09}.

The top panel of Figure~\ref{fig:J1112} shows a portion of a typical light curve for J1112, obtained on 2012 February 1, and includes the brightest comparison star in the field over the same period (the scatter is large because the other comparison stars in the field used to construct this divided light curve are all fainter than $g=16.9$ mag). The bottom panel of this figure shows a Fourier transform (FT) utilizing all $38,863$ light curve points collected thus far. We display the 4$\langle {\rm A}\rangle$ significance line at the bottom of Figure~\ref{fig:J1112}, calculated from the average amplitude, $\langle {\rm A}\rangle$, of an FT within a 1000 \muhz\ region in steps of 200 \muhz, after pre-whitening by the five highest-amplitude periodicities.

\begin{deluxetable}{lcccc}
\tablecolumns{5}
\tablewidth{0.465\textwidth}
\vspace{-0.25in}
\tablecaption{Frequency solution for SDSS J111215.82+111745.0
  \label{tab:J1112freq}}
\tablehead{\colhead{ID} & \colhead{Period} & \colhead{Frequency} & \colhead{Amplitude} & S/N \\
\colhead{} & \colhead{(s)} & \colhead{($\mu$Hz)} & \colhead{(mma)} & \colhead{} }
\startdata
$f_1$ & $2258.528\pm0.003$ & $442.7662\pm0.0007$ & $7.49\pm0.08$ & 26.5 \\
$f_2$ & $2539.695\pm0.005$ & $393.7480\pm0.0007$ & $6.77\pm0.09$ & 23.0 \\
$f_3$ & $1884.599\pm0.004$ & $530.6170\pm0.0011$ & $4.73\pm0.08$ & 16.9 \\
$f_4$ & $2855.728\pm0.010$ & $350.1734\pm0.0013$ & $3.63\pm0.09$ & 11.5 \\
$f_5$ & $1792.905\pm0.005$ & $557.7542\pm0.0017$ & $3.31\pm0.08$ & 11.9 \\ \\

$f_6$ & $134.275\pm0.001$ & $7447.388\pm0.010$ & $0.44\pm0.08$ & {\em 4.4} \\
$f_7$ & $107.56\pm0.04$ & $9297.4\pm3.6$ & $0.38\pm0.14$ & {\em 4.1}
\enddata
\tablecomments{1 mma = 0.1\% relative amplitude}
\end{deluxetable}

The pulse shape of J1112 appears quite sinusoidal, and is nearly solved with five independent periodicities. Those periods have been identified in decreasing order of amplitude in Table~\ref{tab:J1112freq}. For more realistic estimates, the cited errors are not formal least-squares errors to the data but rather the product of $10^5$ Monte Carlo simulations of perturbed data using the software package Period04 \citep{Lenz05}. The signal-to-noise calculation is based on the amplitude of the variability as compared to the average amplitude of a 1000 \muhz\ box centered around that variability, after pre-whitening by the five highest-amplitude periodicities.

A simultaneous linear least-squares fit, fixing these five periods, shows that this variability is quite stable in both amplitude and phase. The $f_1$ periodicity is especially stable in phase, with an r.m.s. scatter less than 7 s between our four months of data---better than 3\% of the $2258.5$ s period. The phase stability of J1112 is reminiscent of hot DAVs such as G117-B15A \citep{Kepler05}, and could be monitored long-term for periodic deviations in arrival times to constrain any possible circumbinary planets \citep{Mullally08}. There is slightly more scatter about the amplitudes measured from month to month, which are more sensitive to variations in photometric conditions. The only periodicity with a consistently decreasing amplitude is $f_2$, which showed an amplitude of 0.7184\% $\pm$ 0.0098\% in 2012 January decrease to 0.573\% $\pm$ 0.016\% in 2012 April.

None of these periods are an integer harmonic of the $4.13952$ hr ($14902.27\pm0.86$ s) orbital period. Given our cited uncertainties, $f_3$, the closest, is more than 8-$\sigma$ from $8 \times f_{orb}$. Thus, tidally induced pulsations cannot properly explain the observed multi-periodic variability. Instead, we conclude that these are global, non-radial $g$-mode pulsations driven to observability by the same mechanism at work in classical DAVs \citep{Brickhill91}. The timescale of this variability is considerably longer than for the pulsations seen in C/O-core DAVs. However, it is consistent with the expectation that the periods of pulsation modes roughly scale with the dynamical timescale for the whole star, $\Pi \propto \rho^{-1/2}$, and are thus much longer for the low-surface-gravity ELM WDs.

\subsection{Potential $p$-mode Pulsations}

\begin{figure}[t]
\centering{\includegraphics[width=0.85\columnwidth]{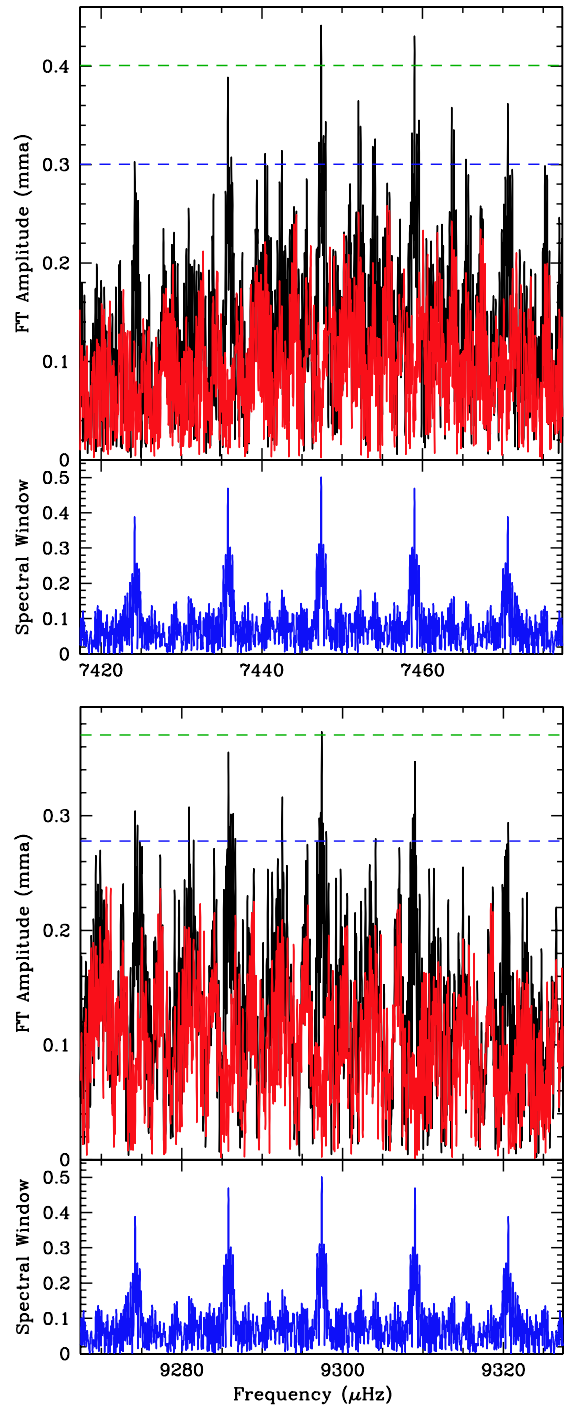}}
\caption{A zoom of the high-frequency regions in the FT of the entire J1112 data set that display evidence for short-period variability, potential $p$-mode pulsations at 134.3 s (top) and 107.6 s (bottom). The original FT is shown in black, and the red shows the residuals after pre-whitening by the highest-amplitude peak. The dashed blue and green lines show the 3$\langle {\rm A}\rangle$ and 4$\langle {\rm A}\rangle$ lines, respectively. The lower panel shows the spectral window in blue centered around each periodicity. \label{fig:J1112pmodes}}
\end{figure}

In addition to the relatively high-amplitude, long-period variability observed in J1112, we see evidence for low-amplitude variability on much shorter timescales. These periodicities are included at the bottom of Table~\ref{tab:J1112freq} in decreasing order of S/N. This S/N value is conservative: We have not pre-whitened by the variability in question for this calculation, which effectively considers some signal as noise in this estimate. We have identified all periodicities with S/N $> 4.0$, and italicize these S/N values to indicate they were calculated in a different way.

We have also computed the probability that each of the short-period detections is real by computing the False Alarm Probability, using the formalism described in \citet{Kepler93}. We find that $f_6$ and $f_7$ have a FAP $> 99.9$\%. (There is an additional peak at 119.552 s with S/N $=3.7$ and a FAP of 99.8\%. However, this periodicity is sufficiently close to the 119.667 s periodic drive error of the 2.1 m Otto Struve telescope that we will not include it in our formal frequency solution.) Figure~\ref{fig:J1112pmodes} shows a zoomed-in portion of the FT around 134.3 s and 107.6 s using our entire data set.

This variability is coherent enough to reach significant amplitude over four months of observations. Some of our longest individual runs also evidence this variability, such as the 6.5 hr run on 2012 Jan 27 (with a $1.24\pm0.36$ mma signal at $134.2\pm2.2$ s) and the 3.5 hr run on 2012 Feb 1 (with a $1.17\pm0.40$ mma signal at $107.9\pm1.8$ s). These two peaks are also fairly significant if we use just our 2012 January data: The 134.275 s mode has $0.55\pm0.17$ mma amplitude ($> 99.9\%$ FAP) while the 107.557 s mode has $0.46\pm0.14$ mma amplitude ($98.2\%$ FAP).

This variability is too short to be explained as $g$-mode pulsations without invoking implausibly high values of the spherical harmonic degree. A recent non-adiabatic pulsation analysis relevant to low-mass WDs by \citet{Corsico12} found that $g$-modes of low radial order (and thus the shortest period) are stable to pulsations and should not be driven to observability. Their calculations found that unstable $g$-modes in a 0.17 \msun\ WD have radial orders $k \geq 9$ and periods longer than 1100 s. Even if we ignore their conclusion that an $\ell=1$, $k=1$ mode is stable, they find this lowest radial order mode has a period $\sim$249.5 s. Similarly, \citet{Steinfadt10} also found that an $\ell=1$, $k=1$ $g$-mode for a 0.17 \msun\ WD has a $\sim$245 s period.

The short-period variability seen in J1112 is also inconsistent with nonlinear combination frequencies present in the non-sinusoidal light curves of many classical DAVs \citep{Brickhill92}. For one, the light curve of J1112 is extremely sinusoidal. Additionally, the short-period variability is not a multiple of any of the five low-frequency modes, nor is it a combination of different modes.

Instead, we propose that this variability is caused by acoustic or pressure ($p$-mode) pulsations driven to observability in J1112. \citet{Corsico12} find that low-order $p$-modes are pulsationally unstable, and have periods ranging from $109-7.5$ s for their $1 < k < 29$ models of a 0.17 \msun\ He-core WD. The 134.3 s period we observe in J1112 is slightly longer than this predicted range, which suggests some uncertainty in identifying the true nature of these instabilities. Still, should these hold up as acoustic modes, this would mark the first detection of $p$-mode pulsations in any WD. We discuss the impact of this discovery in Section~\ref{sec:end}.



\section{SDSS J1518+0658}
\label{sec:J1518}

\subsection{Spectroscopic Observations}

\citet{BrownELMiii} present the spectroscopic discovery data for this $g_0 = 17.5$ mag WD from the Blue Channel spectrograph on the 6.5m MMT. They use 41 separate spectra over more than a year to determine the system parameters, and find that J1518 is in a $14.624\pm0.001$ hr ($52646.4\pm3.6$ s) orbital period binary with a $K=172\pm2$ \kms\ radial velocity semi-amplitude.

We have fit their 41 phased and co-added spectra with the extended stellar atmosphere models of \citet{TB09}, as described in Section~\ref{sec:atm}. This fit formally yields \teff\ $=9900\pm140$ K and \logg\ $=6.80\pm0.05$ for J1518, which corresponds to a mass of $\sim$0.23 \msun\ \citep{Panei07}. Given the mass function ($f = 0.322\pm0.005$ \msun), the minimum mass of the unseen companion is $M_2 > 0.61$ \msun, making it most likely another WD. As with J1112, the nature of the companion has no direct bearing on pulsations in the primary, but it may partially contaminate the spectral fits. Unlike J1112, no metal lines are detected in the spectrum of J1518.

\subsection{Photometric Observations}

\begin{figure}[t]
\centering{\includegraphics[width=0.89\columnwidth]{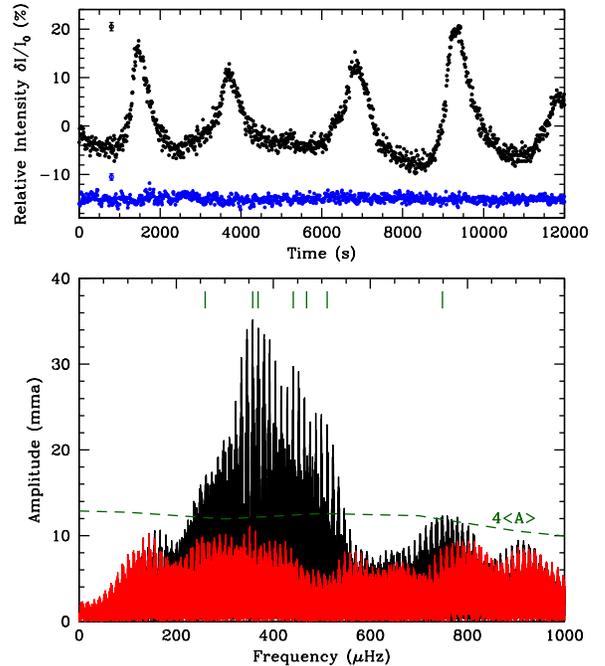}}
\caption{The top panel shows high-speed photometry of J1518 from a representative run, this a portion from 2012 April 16. The brightest comparison star is shown in blue, offset by $-15$\%. Average point-by-point photometric errors are shown. The bottom panel shows a Fourier transform of our entire data set to date, more than 32 hr of observations from 2012~March to 2012~July. We also display in red the Fourier transform of the residuals after pre-whitening by the seven periods listed in the top portion of Table~\ref{tab:J1518freq} and mark those periods with green tick marks at the top of the panel. We mark the 4$\langle {\rm A}\rangle$ significance line as a dashed green line. \label{fig:J1518}}
\end{figure}

\begin{deluxetable}{lllcc}
\tablecolumns{5}
\tablewidth{0.465\textwidth}
\vspace{-0.42in}
\tablecaption{Frequency solutions for SDSS J151826.68+065813.2
  \label{tab:J1518freq}}
\tablehead{\colhead{ID} & \colhead{Period} & \colhead{Frequency} & \colhead{Amplitude} & S/N \\
\colhead{} & \colhead{(s)} & \colhead{($\mu$Hz)} & \colhead{(mma)} & \colhead{} }
\startdata
\multicolumn{5}{c}{\bf Family I} \\
$f_1$ & $2799.087\pm0.005$ & $357.2593\pm0.0007$ & $35.4\pm0.6$ & 11.8 \\
$f_2$ & $2268.203\pm0.004$ & $440.8777\pm0.0007$ & $21.6\pm0.2$ & 7.1 \\
$f_3$ & $2714.306\pm0.003$ & $368.4183\pm0.0005$ & $21.6\pm0.9$ & 7.1 \\
$f_4$ & $1956.361\pm0.003$ & $511.1532\pm0.0007$ & $18.1\pm0.3$ & 5.7 \\
$f_5$ & $3848.201\pm0.009$ & $259.8617\pm0.0006$ & $15.7\pm0.3$ & 5.3 \\
$f_6$ & $1335.318\pm0.003$ & $748.8855\pm0.0015$ & $13.6\pm0.6$ & 4.5 \\
$f_7$ & $2134.027\pm0.004$ & $468.5976\pm0.0008$ & $14.2\pm0.4$ & 4.5 \\
\multicolumn{5}{c}{\bf Family II} \\
$f_1$ & $2796.048\pm0.022$ & $357.6476\pm0.0028$ & $41.1\pm4.8$ & 13.5 \\
$f_2$ & $2209.996\pm0.024$ & $452.4895\pm0.0049$ & $19.9\pm1.4$ & 6.5 \\
$f_3$ & $2802.769\pm2.6$   & $356.79\pm0.33$     & $26.4\pm3.9$ & 8.7 \\
$f_4$ & $1956.331\pm0.008$ & $511.1609\pm0.0020$ & $18.1\pm0.8$ & 5.7 \\
$f_5$ & $3683.703\pm7.7$   & $271.47\pm0.57$     & $17.7\pm3.2$ & 5.7 \\
$f_6$ & $2413.091\pm29.5$  & $414.41\pm5.07$     & $15.6\pm4.0$ & 5.2 \\
$f_7$ & $1318.847\pm0.004$ & $758.2382\pm0.0021$ & $12.9\pm0.4$ & 4.4
\enddata
\end{deluxetable}

We obtained photometric observations of J1518 from 2012 March to 2012 July for more than 32.2 hr of coverage; a full journal of observations can be found in Table~\ref{tab:jour}. The data were obtained and reduced in an identical manner to those of J1112, described in Section~\ref{sec:J1112photo}. We divided the sky-subtracted light curves by the two brightest comparison stars in the field, SDSS~J151824.11+065723.2 ($g=14.0$ mag) and SDSS~J151828.78+065928.9 ($g=14.3$ mag), to allow for fluctuations in seeing and cloud cover.

The top panel of Figure~\ref{fig:J1518} shows a portion of a typical light curve for J1518, obtained in 2012 April, and includes the brightest comparison star in the field over the same period. High-amplitude, multi-periodic, highly non-sinusoidal variability is evident. The bottom panel of this figure shows an FT utilizing all $16,522$ light curve points collected thus far.

Given the relatively sparse coverage over four months ($<1.2$\% duty cycle), the spectral window for our observations of J1518 is quite messy, and aliasing from gaps in the data makes identifying the underlying periods of variability especially difficult. As such, we have provided two families of solutions in Table~\ref{tab:J1518freq}.

The highest peak in the FT occurs at $\sim$357 \muhz. However, after pre-whitening by this frequency, there is considerable ambiguity as to the next highest peak. There are two peaks with nearly the same amplitude, one slightly higher at 452.4895 \muhz\ than the other split apart by the daily alias at 440.8777 \muhz. Our two families of solutions are based on which of these aliases we pick. We note that the top family of solutions holds up more robustly to a Monte Carlo simulation of the associated frequency and amplitude uncertainties, but the bottom family does a marginally better job of reproducing the data (the residuals after fitting this solution have a 3.27\% total amplitude, while the residuals after fitting the top family have 3.32\% total amplitude). In both cases, the fits do a poor job of reproducing the non-linear peakiness manifest as the sharp rises and falls in the light curve. However, the fits do a decent job of predicting when these features occur.

The seven significant periodicities we identify in J1518 are by no means exhaustive; we note the significant residual power left over after pre-whitening by these periods shown as the red FT at the bottom of Figure~\ref{fig:J1518}. In the range between $200-800$ \muhz\ ($1250-5000$ s), the average amplitude of the original FT is 6.4 mma, and after pre-whitening it remains above 3.4 mma. A coordinated campaign of near-continuous observations of this relatively bright WD would greatly help resolve the periods present in this pulsating WD.



\section{Discussion}
\label{sec:end}

\subsection{Acoustic ($p$-mode) Pulsations in WDs}

As with the C/O-core DAVs that have been known for more than 40 years, the dominant optical variability in these ELM WDs is consistent with surface temperature variations caused by non-radial $g$-mode pulsations driven to observability by a hydrogen partial ionization zone. However, in J1112, we also see the first evidence for short-period $p$-mode pulsations in a WD. These $p$-modes offer a tantalizing opportunity to probe the interior of an ELM WD in a complimentary way to the $g$-modes also present.

Pulsation calculations have long shown that WDs of all masses should be unstable to $p$-mode pulsations (e.g. \citealt{Saio83,Starrfield83,Hansen85}). However, despite exhaustive searches (e.g. \citealt{Robinson84,Kawaler94,Silvotti11}), no such high-frequency modes have ever been observed in a WD. Discovery of $p$-mode pulsations in ELM WDs could suggest that these pulsation models are indeed valid, but that the amplitudes of these oscillations are simply too small to detect with significance in the more massive DAVs.

In the context of hot B subdwarf stars (sdBs), it is not entirely surprising to find ELM WDs with observable $p$-mode pulsations. Variable hot B subdwarf stars (sdBVs), especially of the EC 14026 or V361 Hya class, are observed to vary with $p$-mode pulsations (e.g. \citealt{Kilkenny97,Charpinet97}). These sdBVs are qualitatively quite similar to ELM WDs, with He-cores and similar surface gravities, $5.2 <$ \logg\ $< 6.1$, although they are typically much hotter, with \teff\ $>$ 28,000 K, and more massive, with canonical masses $\sim$0.45 \msun. Their pulsation periods are most often in the range $120-300$ s \citep{Kilkenny07}, quite similar to the variability in J1112. Many sdBVs are hybrid pulsators, showing both $p$- and $g$-mode pulsations.

Still, we must be careful to rule out all other possibilities before declaring the short-period instabilities in J1112 as bona-fide $p$-modes. The low-order $p$-modes calculated by \citet{Corsico12} range from $109-7.5$ s for their $1 < k < 29$ models of a 0.17 \msun\ He-core WD, slightly shorter than the 134.3 s periodicity in J1112. However, since that work finds that $g$-modes with periods shorter than 1100 s are stable to pulsations, $p$-mode pulsations remain the most likely explanation for the short-period variability in J1112. (It also remains possible that the 134.3 s periodicity is a radial fundamental mode for the star.)

\begin{figure*}[t]
\centering{\includegraphics[width=0.65\textwidth]{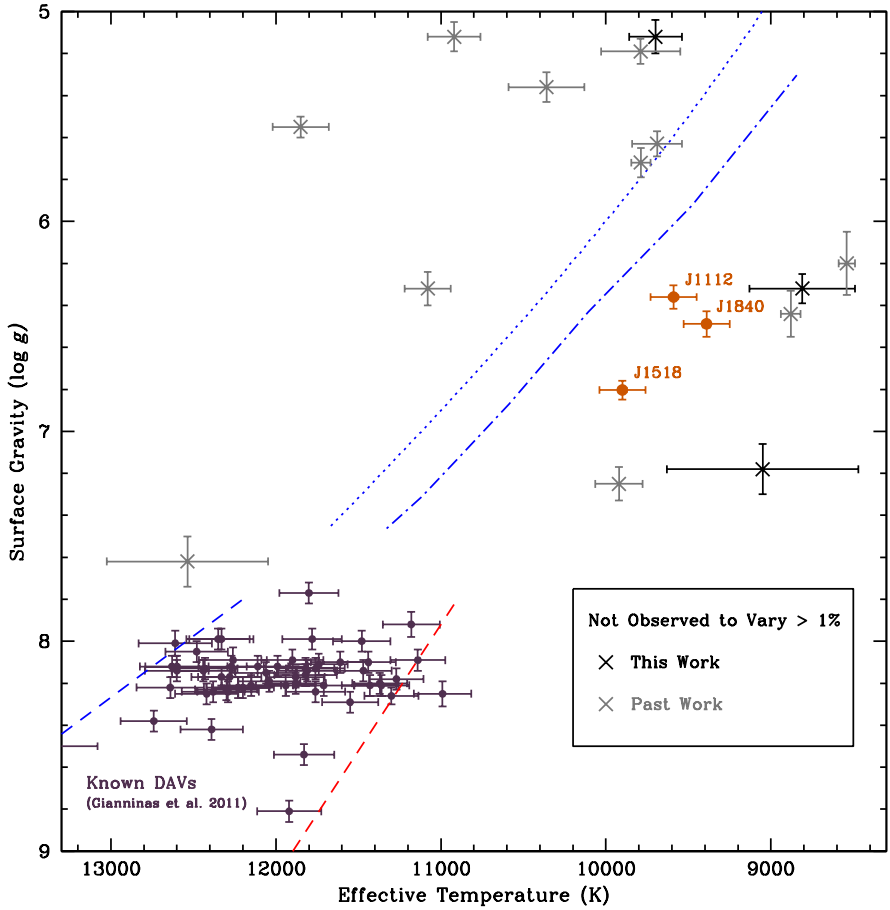}}
\caption{The extended ZZ Ceti instability strip. The 56 previously known C/O-core DAVs characterized by \citet{Gianninas11} are included as purple dots, and our new pulsating ELM WDs are marked in burnt orange. We denote an extrapolated theoretical blue edge for the low-mass ZZ instability strip; this dotted blue line is described in the text and uses the criterion $2 \pi \tau_C = 100$ s, with the convective prescription ML2/$\alpha$=1.5. We include the theoretical blue edge for low-mass DAVs from \citet{VanGrootel13} as a dashed-dotted blue line. We also mark the empirical blue- and red-edges from \citet{Gianninas11} as dashed lines. Objects not observed to vary to at least greater than 10 mma (1\%) are marked with Xs. We include three new WDs not observed to vary, listed in Table~\ref{tab:null}; the others from past work, in gray, were detailed in \citet{HermesJ1840} and \citet{Steinfadt12}. \label{fig:search}}
\end{figure*}

\subsection{Asteroseismic Applications}

While we have not yet matched the observed periods in these first three pulsating ELM WDs to full asteroseismic models, we can begin to assess some of their observed properties. We are extremely interested in finding pulsation periods that are consecutive radial overtones, since the asymptotic period spacings of $g$-modes are a direct probe of the overall mass of the star. We expect that low-mass pulsating WDs will have relatively high mean period spacings. For the relevant He-core WD models, \citet{Corsico12} find that 0.17 \msun\ WDs have dipole forward period spacings of roughly 104 s, compared to period spacings of 64 s for a 0.45~\msun\ He-core WD. Similarly, \citet{Steinfadt10} find an 89 s mean period spacing for a 0.17 \msun\ WD.

The rich and relatively simple pulsation spectrum of J1112 offers an excellent opportunity to search for consecutive radial overtones. The two shortest-period $g$-mode pulsations in this star, at 1884.6 s and 1792.9 s, differ by 91.7 s. If these are $\ell=1$ modes of consecutive radial order, this would be excellent direct confirmation that this is indeed a roughly 0.17~\msun\ He-core WD. However, a full asteroseismic investigation is warranted to confirm that these are indeed consecutive modes. That hinges on constructing enough He-core WD models with sufficiently different hydrogen layer masses.

Additionally, J1112 (relatively sinusoidal) and J1518 (extremely nonlinear) offer strongly contrasting pulse shapes. It has been suggested that the large nonlinear distortions in the emergent flux are due to the changing depth of the star's convection zone during a pulsation cycle \citep{Brickhill92,Wu01,Montgomery05}. If this is the case, the method of \citet{Montgomery10} could be used on J1518 to infer the thermal response timescale and thereby the average depth of the convection zone in this highly non-sinusoidal DAV. Distortions in the light curve may also be the result of other processes, such as the nonlinear response of the emergent flux to temperature perturbations \citep{FontBrass08}.

Finally, we can put all three pulsating ELM WDs into context by calculating their weighted mean periods (WMPs), as defined by \citet{Mukadam06}. In the classical C/O-core ZZ Ceti stars, cooler WDs typically have longer WMPs. This generally holds for the pulsating ELM WDs, as well: The coolest of the three, J1840, has a WMP $\sim$3722 s. J1112 and J1518 have WMPs $\sim$2288 s and $\sim$2404 s, respectively. By comparison, no classical C/O-core DAV known before the discovery of pulsating ELM WDs had a WMP $> 1200$ s (or \teff\ $<$ 10,000 K).

\begin{deluxetable}{llll}
\tablecolumns{4}
\tablewidth{0.465\textwidth}
\vspace{-0.225in}
\tablecaption{Properties of the Three Known Pulsating ELM WDs
  \label{tab:3props}}
\tablehead{\colhead{Property} & \colhead{Value} & \colhead{Property} & \colhead{Value} }
\startdata
\multicolumn{4}{c}{\bf SDSS~J184037.78+642312.3} \\
\teff\  & $9390\pm140$ K    				& \logg\          & $6.49\pm0.06$ \\
Mass    & $\sim$0.17 \msun\ 				& $g_0$ & 18.8 mag \\
Periods & $2094-4890$ s     				& Max Amp.        & $>5.1$\% \\
$P_{{\rm orb}}$ & $4.5912\pm0.001$ hr 	& $M_2$ & $>0.64$ \msun\ \\
\multicolumn{4}{c}{\bf SDSS~J111215.82+111745.0} \\
\teff\  & $9590\pm140$ K    				& \logg\          & $6.36\pm0.06$ \\
Mass    & $\sim$0.17 \msun\ 				& $g_0$ & 16.2 mag \\
Periods & $107.6-2855$ s    				& Max Amp.        & $>0.7$\% \\
$P_{{\rm orb}}$ & $4.1395\pm0.0002$ hr	& $M_2$ & $>0.14$ \msun\ \\
\multicolumn{4}{c}{\bf SDSS~J151826.68+065813.2} \\
\teff\  & $9900\pm140$ K    				& \logg\          & $6.80\pm0.05$ \\
Mass    & $\sim$0.23 \msun\ 				& $g_0$ & 17.5 mag \\
Periods & $1335-3848$ s     				& Max Amp.        & $>3.5$\% \\
$P_{{\rm orb}}$ & $14.624\pm0.001$ hr  	& $M_2$ & $>0.61$ \msun\
\enddata
\end{deluxetable}

These pulsating ELM WDs lie in a new region of the classical ZZ Ceti instability strip, extending the strip to much cooler and lower surface gravity WDs. We have used our new temperature and surface gravity determinations described in Section~\ref{sec:atm} to characterize our three new ELM WDs. This means we have also updated the atmospheric parameters for J1840, the first pulsating ELM WD discovered. We also outline the new period determinations from J1840, as detailed in \citet{Corsico12}. These values can be found in Table~\ref{tab:3props}.

\begin{deluxetable*}{lcccrc}
\tablecolumns{6}
\tabletypesize{\footnotesize}
\tablewidth{1.0\textwidth}
\vspace{-0.35in}
\tablecaption{Observed Low-Mass DAV Candidates and Null Results\label{tab:null}}
\tablehead{\colhead{Object} & \colhead{$g_0$-SDSS} & \colhead{\teff} &
 \colhead{\logg} & \colhead{Reference} & \colhead{Det. Limit} \\ 
 \colhead{} & \colhead{(mag)} & \colhead{(K)} & \colhead{(cm s$^{-1}$)} & \colhead{} & \colhead{\%} }
\startdata
SDSS~J012549.37+461920.1 & 15.8 & $9050\pm580$  & $7.18\pm0.12$ & \citet{BrownELMiii} & 0.3 \\
SDSS~J144342.74+150938.6 & 18.6 & $8810\pm320$  & $6.32\pm0.07$ & \citet{BrownELMiii} & 0.1 \\
SDSS~J221928.48+120418.6 & 17.7 & $9700\pm160$  & $5.12\pm0.08$ & \citet{BrownELMiii} & 0.3
\enddata
\end{deluxetable*}

We continue to search for new pulsating ELM WDs, but there is also utility in constraining the regions of parameter space where WDs are not observed to vary. In addition to the targets shown not to vary by \citet{HermesJ1840}, we have put limits on the lack of variability in three new ELM WDs, detailed in Table~\ref{tab:null}. Our limits on SDSS~J144342.74+150938.6 are particularly stringent since we observed its field for more than 7.2 hr; one of the comparison stars we used, SDSS~J144347.31+150841.9, turned out to be variable, most likely a $\delta$-Scuti star, dominated by a 3.46-hr pulsation period. We construct an updated empirical instability strip using the three new pulsators and more than a dozen low-mass WDs not observed to vary, shown in Figure~\ref{fig:search}.

The addition of two new pulsating ELM WDs aids in constraining the theoretical low-mass WD instability strip. Following \citet{Brickhill91} and \citet{Goldreich99}, we use the criterion that $P_{\rm max} \sim 2 \pi \tau_{\rm C}$ for the longest period mode that is excited, where $P_{\rm max}$ is the mode period and the timescale $\tau_{\rm C}$ describes the heat capacity of the convection zone as a function of the local photospheric flux, which we compute from a grid of models\footnote{Note that $\tau_{\rm C} \approx 4 \tau_{\rm th}$, where $\tau_{\rm th}$ is the thermal timescale at the base of the convection zone. The average value of $\tau_C$ is often denoted as $\tau_0$.}. Using the traditional value of $P_{\rm max} = 100\,$s, we find that the convection parameters ML2/$\alpha=1.5$ provide a good match to the blue edge for the normal-mass DAVs, but that extrapolating this relation to the pulsating ELM WDs yields a blue edge which is somewhat hotter than the observations (see the dotted blue line in Figure~\ref{fig:search}). The models of \citet{Corsico12} find that only $g$-mode periods greater than $\sim$1000 s are excited, but using this criterion only moves the blue edge a small amount closer to the location of the observed pulsators (about 130 K for a given surface gravity). It is possible there are additional mechanisms at work that move the blue edge to cooler temperatures. Using a less efficient prescription for convection, such as ML2/$\alpha=1.0$, will shift the blue edge to cooler temperatures, as seen in \citealt{VanGrootel13}, whose blue edge we include in Figure~\ref{fig:search}. We are also still dominated by small number statistics, and perhaps we have simply not yet found any hotter pulsating low-mass WDs. It remains difficult to predict a theoretical red edge for even the classical C/O-core DAVs, so we have not extended it to low-mass WDs, and instead provide a portion of the red-edge established by \citet{Gianninas11}.

\section{Conclusions}

We have discovered the second and third pulsating ELM WDs as part of a search for such objects at the McDonald Observatory. Spectroscopic fits to both objects indicate each has a mass $<$ 0.25 \msun, and thus both likely harbor a helium core. These variable WDs are most likely members of a low-mass extension of the classic ZZ Ceti instability strip.

We will continue our efforts to populate the empirical low-mass WD instability strip, and look forward to the exciting prospect of performing asteroseismology by matching the observed pulsation periods in low-mass WDs to theoretical He-core WD models.

\acknowledgments

We acknowledge Gilles Fontaine and Lars Bildsten for extremely helpful comments. The authors are especially grateful to the students of the UT Freshmen Research Initiative who assisted with this data acquisition and reduction, including John Pelletier, Arina Rostopchina, Aivien Do, Oscar Roussett, and Luis Castro. J.J.H., M.H.M. and D.E.W. acknowledge the support of the NSF under grant AST-0909107 and the Norman Hackerman Advanced Research Program under grant 003658-0252-2009. The authors are grateful to the tireless assistance of the McDonald Observatory support staff, and to Fergal Mullally for developing some of the data analysis pipeline used here.

\end{document}